\documentclass[12pt]{iopart}
\usepackage{iopams}  
\usepackage{setstack}
\usepackage{bm}
\usepackage[breaklinks=false,colorlinks,citecolor=blue,linkcolor=blue,urlcolor=blue]{hyperref}
\begin{document}

\title{Discrete-time construction of nonequilibrium path integrals on the Kostantinov-Perel' time contour}

\author{Andrea Secchi, Marco Polini}

\address{Istituto Italiano di Tecnologia, Graphene Labs, Via Morego 30, I-16163 Genova,~Italy}
\ead{andrea.secchi@iit.it}
\vspace{10pt}
\begin{indented}
\item[]March 2019
\end{indented}

\begin{abstract}
Rigorous nonequilibrium actions for the many-body problem are usually derived by means of path integrals combined with a discrete temporal mesh on the Schwinger-Keldysh time contour. The latter suffers from a fundamental limitation: the initial state on this contour cannot be arbitrary, but necessarily needs to be described by a non-interacting density matrix, while interactions are switched on adiabatically. The Kostantinov-Perel' contour overcomes these and other limitations, allowing generic initial-state preparations. In this Article, we apply the technique of the discrete temporal mesh to rigorously build the nonequilibrium path integral on the Kostantinov-Perel' time contour.
\end{abstract}

%
\vspace{2pc}

\section{Introduction}

The theory of many-body systems brought away from equilibrium strongly relies on the concept of time contours. Introduced in different forms by several authors, including Schwinger~\cite{Schwinger}, Keldysh~\cite{Keldysh}, and Kostantinov and Perel'~\cite{KostantinovPerel}, time contours provide an elegant way to deal on equal footing with time-ordered and anti-time-ordered products of operators which are both needed in the calculation of time-dependent observables out of equilibrium. For appropriately defined time contours, such cumbersome machinery is replaced by a single time-ordering procedure. The price to pay is that the size of the time domain has to be doubled with respect to the physical real-time axis. 

Typically, one uses a time contour to evaluate nonequilibrium Green's functions and associated observables. On both the Schwinger-Keldysh and the Kostantinov-Perel' contours, equations of motion can be written down straightforwardly and many-body perturbation theory can be developed~\cite{Keldysh, KadanoffBaym, RammerSmith, Stefanucci, vanLeeuwen, Odashima17, Kantorovich18}. In particular, the Kadanoff-Baym equations~\cite{KadanoffBaym} are the cornerstone of any numerical calculation of nonequilibrium Green's functions for correlated systems~\cite{Bonitz13, Schuler18, Karlsson18, Abdurazakov18, Hopjan19}.

Complementary to numerical approaches, path integrals constitute, in the context of nonequilibrium physics, a very useful analytical technique to derive effective theories, kinetic equations, and to define transparent physical approximations~\cite{Kamenev, Taniguchi17, deNicola19}. These derivations typically rely on nonequilibrium generalizations of procedures which appear also with equilibrium path integrals, such as the Gaussian integration over a subspace of fields (e.g.~integration over bosons in a coupled fermion-boson system), the Hubbard-Stratonovich transformation, and the saddle-point approximation. 

For the path integral to be meaningful, the action should be defined in a rigorous way, in particular, taking care of the correct operational definition of the inverse single-particle Green's function operator, whose form depends on the time contour. Since the time contours used for nonequilibrium theories are different than the equilibrium (Matsubara) one, it is necessary to set up the corresponding actions case by case. This has been done for the Schwinger-Keldysh (SK) time contour~\cite{Kamenev}. Here, we argue that a similarly rigorous derivation for the case of the Kostantinov-Perel' (KP) time contour is timely. 

In what follows, we first summarize the main features of SK and KP time contours, respectively. We denote by $\gamma_{\rm K} = \gamma_{{\rm K}, +} \cup \gamma_{{\rm K}, -}$ the SK contour. Here $\gamma_{{\rm K}, +}$ and $\gamma_{{\rm K}, -}$ are the forward and backward branches of the real-time axis $(-\infty, + \infty)$, respectively. These are obtained by doubling the real-time degrees of freedom: $\gamma_{{\rm K}, +}$ consists of the real axis traveled forward from $- \infty$ to $+ \infty$, while $\gamma_{{\rm K}, -}$ consists of the real axis traveled backwards from $+ \infty$ to $- \infty$. Nonequilibrium theories on the SK contour require to specify the density matrix, $\hat{\rho}_{- \infty}$, which describes the system in the remote past, i.e.~for $t \rightarrow - \infty$. Crucially, this density matrix is required to be non-interacting~\cite{Kamenev}. Many-body interactions, if present, are switched on adiabatically in such a way that the Hamiltonian coincides with the physical one at a given time $t = t_{0}$. 

The KP contour~\cite{KostantinovPerel, Stefanucci} allows a much greater flexibility with respect to the choice of the initial state of the system. In this formalism, one specifies the preparation of the system at an arbitrary time $t_{0}$ ($> - \infty$) via the density matrix $\hat{\rho}_{0}$, whose most general form is~\cite{Stefanucci}
\begin{eqnarray}
\hat{\rho}_0 \equiv   \mathrm{e}^{- \beta \hat{ \cal H}_{\rm M} }   \biggl/  \mathrm{Tr}(\mathrm{e}^{- \beta \hat{ \cal H}_{\rm M} })  ~,
\label{rho_0}
\end{eqnarray}
where $\beta > 0$ is a positive constant and $\hat{ \cal H}_{\rm M}$ is an {\it arbitrary} Hermitian operator. A standard ({\it but not mandatory}) choice is 
\begin{eqnarray}
\hat{ \cal H}_{\rm M} = \hat{\cal H}(t_0) - \mu \hat{\cal N} ~,
\label{standard HM}
\end{eqnarray}
where $\hat{\cal H}(t_0)$ is the physical Hamiltonian of the system at time $t = t_0$, $\mu$ is the chemical potential, and $\hat{\cal N}$ is the number operator. This describes a system at thermal equilibrium at the initial time, in the grand-canonical ensemble, with inverse temperature $\beta$. However, it should be emphasized that $\hat{ \cal H}_{\rm M}$ can be chosen to be a much more general operator~\cite{Stefanucci}, which allows to select a specific initial state, avoiding to assign the same weight to degenerate eigenstates of the Hamiltonian $\hat{\cal H}(t_0)$. This allows to go beyond the grand-canonical ensemble and, for example, to select a broken-symmetry state instead of a thermal mixture, something which cannot be captured with the choice in equation~(\ref{standard HM}). (We will show this by means of a concrete example in section~\ref{sect:example}.) Importantly, the KP formulation also allows to consider fully interacting systems without resorting to the adiabatic switching on of interactions, and is therefore appropriate for the study of strongly correlated systems. Since the adiabatic switching on is bypassed, one can control the preparation of the system at a finite time $t_{0}$, rather than in the far past. Because of this flexibility, the KP contour is a suitable framework for theories involving coupled fermion-boson systems, such as in the problem of nonequilibrium superconductivity~\cite{Secchi17} and/or initial broken-symmetry states, such as in nonequilibrium magnetism~\cite{Secchi16PRB}.
 
In the latter formalism, the real-time domain is $[ t_0,  \infty)$, and the KP time contour $\gamma$ is given by the union of three branches: $\gamma = \gamma_+ \cup \gamma_- \cup \gamma_{\rm M}$. The forward ($\gamma_+$) and backward ($\gamma_-$) branches are analogous to the SK branches, except that their domain of definition is $[t_{0},  \infty)$, while the Matsubara branch ($\gamma_{\rm M}$) is the imaginary-time domain needed to describe the initial statistical mixture, $[t_0, t_0 - i \beta)$, where $\beta$ is the initial inverse temperature ($\hbar = 1$ throughout this Article). One can write equation~(\ref{rho_0}) in terms of an evolution operator along the Matsubara branch, i.e.,
\begin{eqnarray}
\hat{\rho}_0 \equiv     \hat{ \cal U}_{{\gamma_{\rm M}}}    \biggl/ \mathrm{Tr}( \hat{ \cal U}_{{\gamma_{\rm M}}}) ~.
\end{eqnarray}  

While the Green's function problem on the KP contour is thoroughly  discussed in the literature~\cite{Stefanucci}, a rigorous path-integral formulation seems to be available only in the case of the SK contour~\cite{Kamenev}. Given the usefulness of path integrals when dealing with problems that can be simplified by field integration~\cite{Secchi17}, it is convenient to set up the same tools for the KP contour. The problem basically consists in deriving the KP nonequilibrium action, which correctly keeps into account the {\it boundary terms} arising from the construction of the path integral. This is the goal of the present Article.

Our Article is organized as follows. In section~\ref{sect:Hamiltonian} we introduce a generic time-dependent Hamiltonian, including both fermionic and bosonic degrees of freedom. In section~\ref{sect:pathint} we present the nonequilibrium partition function and the associated action that we need to derive from ``first principles''. In section~\ref{sect: KP action} we apply the technique of the discrete temporal mesh to derive the discrete form of the action. In section~\ref{sect: ff fb Gf} we derive the general form of the free-fermion and free-boson Green's functions on the KP contour, which allows to treat  boundary conditions exactly while moving to the continuous time representation. We then show several important simplifications that occur in relevant particular cases. In section~\ref{sect: discrete continuum} we complete the transition to the continuous-time representation of the action. In section~\ref{sect:example} we present the solution of a simple nonequilibrium model using both ordinary quantum mechanics and the KP path integral, with the aim of demonstrating the flexibility of the latter in fixing the initial conditions. Finally, in section~\ref{sect:summary} we summarize our main results and their applicability.

\section{Hamiltonian}
\label{sect:Hamiltonian}

We consider a general system of fermions and bosons described by the following time-dependent Hamiltonian:
\begin{eqnarray}
\hat{\cal H}(t) \equiv \hat{\cal H}_{\rm f}(t) + \hat{\cal H}_{\rm b}(t) +  \hat{\cal I}(t)~.
\label{Hamiltonian}
\end{eqnarray}
Here,
\begin{eqnarray}\label{free-el Hamiltonian Nambu}
\hat{\cal H}_{\rm f }(t) =   \hat{d}^{\dagger} \cdot \mathbf{T}_{\rm f }(t) \, \hat{d}  
\end{eqnarray}
is the single-fermion Hamiltonian,
\begin{eqnarray}\label{free-ph Hamiltonian Nambu}
\hat{\cal H}_{\rm b}(t) =      \hat{a}^{\dagger} \cdot \mathbf{T}_{\rm b }(t) \, \hat{a} 
\end{eqnarray}
is the single-boson Hamiltonian, and $\hat{\cal I}(t)$ includes all the remaining terms. The creation/annihilation operators, $\hat{d}^{\dagger}$/$\hat{d}$ for fermions and $\hat{a}^{\dagger}$/$\hat{a}$ for bosons, respectively, are grouped into vectors whose components are distinguished by single-particle quantum numbers (e.g.~wavevector and spin projection for electrons on a lattice). We do not make any assumption on the physical nature of the fields that were just introduced. Fermions can be e.g.~electrons, holes, or Nambu fermions. Bosons can be e.g.~photons or phonons. The term $\hat{\cal I}(t)$ may include any form of interaction between these fields, that is, both between particles of the same species (e.g.~electron-electron Coulomb interaction) and between particles of different species (e.g.~electron-phonon coupling). The form of the interaction terms does not affect the derivation of the single-particle inverse Green's function operator, which is the main mathematical issue that we solve here. The single-particle Hamiltonian matrix $\mathbf{T}_{\mathrm{f/b}}(t)$ can include a time-independent hopping (or, if the hopping is diagonalized, the single-particle energy spectrum), as well as the effect of any external time-dependent fields. Setting either $\mathbf{T}_{\mathrm{f}}(t) = 0$ in~(\ref{free-el Hamiltonian Nambu}) or $\mathbf{T}_{\mathrm{b}}(t) = 0$ in~(\ref{free-ph Hamiltonian Nambu}) allows one to restrict the theory to the particular cases of purely bosonic or fermionic systems, respectively. In the following, we will always use bold fonts to denote matrices in the basis of single-particle quantum numbers.

\section{Nonequilibrium path integrals}
\label{sect:pathint}

We denote the contour variable by $z \in \gamma$. For every real time coordinate $t \in [t_0, \infty)$, there are two values of $z$: one belonging to $\gamma_+$, which we denote by $t_+$, and one belonging to $\gamma_-$, which we denote by $t_-$. For every imaginary-time coordinate on the Matsubara branch $[t_0, t_0 - i \beta)$ there is a single value of $z$.

The nonequilibrium partition function is expressed as a coherent-state path integral over Grassmann variables $(\overline{d}, d)$ (representing the fermion fields) and complex variables $(a^*, a)$ (representing the boson fields)~\cite{Kamenev}. As in the previous section, with a single Latin letter, such as ``$d$'', we intend an array of variables, one for each different value of the set of single-particle quantum numbers. In the coherent-state representation, these variables acquire a further dependence on the contour variable $z$. The general expression for the partition function is
\begin{eqnarray}\label{ZKB}
Z\left[ V \right] \equiv \frac{\mathrm{Tr}\left\{\hat{\mathcal{U}}^{(V)}_{\gamma}\right\}}{\mathrm{Tr}( \hat{ \cal U}_{{\gamma_{\rm M}}}) }  
     \equiv \frac{1}{\mathrm{Tr}( \hat{ \cal U}_{{\gamma_{\rm M}}})} \int \mathcal{D} \left(\overline{d}, d \right) \int \mathcal{D} \left( a^*, a \right) \mathrm{e}^{i S^{(V)}\left[\overline{d}, d ; a^*, a \right]}~,
\end{eqnarray}
where $\hat{\mathcal{U}}^{(V)}_{\gamma} $ is the evolution operator along the contour $\gamma$ , namely $\hat{\mathcal{U}}^{(V)}_{\gamma} = \hat{\mathcal{U}}_{\gamma_{\rm M}} \hat{\mathcal{U}}^{(V)}_{\gamma_-} \hat{\mathcal{U}}^{(V)}_{\gamma_+}$ and we have included some source potential $V(z)$ depending on the contour variable $z$. One has $\hat{\mathcal{U}}^{(0)}_{\gamma}  = \hat{\mathcal{U}}_{\gamma_{\rm M}}$ and therefore $Z[V = 0] = 1$. The same identity, i.e.~$Z = 1$, applies if $V(t_{+}) = V(t_{-})$, $\forall t$. For simplicity, we assume that the sources are quadratic in the particle fields. The nonequilibrium action is
\begin{eqnarray}
S^{(V)}[\overline{d}, d ; a^*, a]   \equiv  S^{(V)}_{\mathrm{f}, Q}[\overline{d}, d] + S^{(V)}_{\mathrm{b}, Q}[a^*, a]     +  S_{\cal I}[\overline{d}, d ; a^*, a]~,
\label{action KB}
\end{eqnarray}
where 
\begin{eqnarray}
S_{\mathrm{f}, Q}^{(V)}[\overline{d}, d] \equiv  \int_{\gamma} \mathrm{d} z\Big\{ \overline{d}(z)  \cdot i \partial_z   d(z)  -    {\cal H}^{(V)}_{\rm f}[\overline{d}(z), d(z)  ; z ] \Big\} 
\label{S_f}
\end{eqnarray}
and
\begin{eqnarray}
 S_{\mathrm{b}, Q}^{(V)}[a^*, a]  \equiv \! \int_{\gamma} \mathrm{d} z \Big\{  a^*(z)\!  \cdot \! i \partial_z   a(z)  -    {\cal H}^{(V)}_{\mathrm{b}}[ a^*(z), a(z)   ; z ]\Big\}  
 \label{S_b} 
\end{eqnarray}
contain all the terms which are quadratic in the fermion or boson fields, respectively, while
\begin{eqnarray}
 S_{\cal I}\left[\overline{d}, d ; a^*, a \right] \equiv  -\int_{\gamma} \mathrm{d} z  \, {\cal I}\left[\overline{d}(z), d(z) ; a^*(z), a(z); z \right]   
 \label{S_IV} 
\end{eqnarray}
includes all the other terms, such as interactions between particles of the same or different kind. It is important to notice that, differently from the case of the SK contour, in the present case we must include the Matsubara branch and account for the fact that the Matsubara Hamiltonian $\hat{\cal{H}}_{\rm M}$, introduced in~(\ref{rho_0}), is, in general, different from the physical Hamiltonian at the initial time, $\hat{\cal H}(t_0)$. In general, one has
\begin{eqnarray}
\hat{\cal{H}}_{\rm M} = \hat{\cal H}(t_0) - \mu_{\rm f} \hat{\cal N}_{\rm f} - \mu_{\rm b} \hat{\cal N}_{\rm b}    + \hat{\cal R}~,
\end{eqnarray}
where $\mu_{\rm f ( \rm b )}$ is the chemical potential for the fermion (boson) subsystem, $\hat{\cal N}_{\rm f (\rm b)}$ is the fermion (boson) number operator, and $\hat{\cal R}$ includes all the remaining fermionic and bosonic terms. If the equilibrium state is a grand-canonical thermal distribution, then $\hat{\cal R} = 0$. If, instead, one wants to implement a non-thermal initial state~\cite{Stefanucci}, a non-vanishing $\hat{\cal R}$ is needed.

As in the previous discussion about the physical Hamiltonian, we separate the terms of the Matsubara Hamiltonian which are quadratic in the particle fields from those which are not. The quadratic terms have the form $\hat{d}^{\dagger} \cdot \mathbf{T}^{\mathrm{M}}_{\mathrm{f}} \,  \hat{d}$ and $\hat{a}^{\dagger} \cdot \mathbf{T}^{\mathrm{M}}_{\mathrm{b}} \,  \hat{a}$, where we have introduced the Matsubara hopping matrices $\mathbf{T}^{\mathrm{M}}_{\mathrm{f/b}}$. After moving to the path integral formulation, all the terms which are not quadratic are collected into the quantity ${\cal I}\left[\overline{d}(z), d(z) ; a^*(z), a(z); z \right]$, introduced in equation~(\ref{S_IV}), while the quadratic terms are included into equations~(\ref{S_f}) and (\ref{S_b}) as
\begin{eqnarray}
{\cal H}^{(V)}_{\rm f}[\overline{d}(z), d(z)  ; z] \equiv   \overline{d}(z) \cdot \mathbf{T}^{(V)}_{\mathrm{f}}(z) \,  d(z)
\label{contour Hf}
\end{eqnarray}
and
\begin{eqnarray}
{\cal H}^{(V)}_{\rm b}[a^*(z), a(z)  ; z] \equiv  a^*(z) \cdot \mathbf{T}^{(V)}_{\mathrm{b}}(z) \,  a(z)~,
\label{contour Hb}
\end{eqnarray}
where
\begin{eqnarray}
\mathbf{T}^{(V)}_{\mathrm{f/b}}(z)    \equiv \Theta(t_{0-}, z)  \left[ \mathbf{T}_{\mathrm{f/b}}(t) + \mathbf{V}_{\mathrm{f/b}}(z) \right]      + \Theta(z, t_{0 -}) \, \mathbf{T}^{\mathrm{M}}_{\mathrm{f/b}} ~.   
\label{contour hopping}
\end{eqnarray}
In the above equations, $\Theta(z, t_{0 -})$ equals $1$ if $z$ lies on the Matsubara branch, and $0$ otherwise ($t_{0-}$ is the initial time taken on the backward branch). On the other hand, $\Theta( t_{0 -}, z)$ equals $1$ if $z$ lies on either $\gamma_+$ or $\gamma_-$, and $0$ otherwise. We have also included the quadratic sources into~(\ref{contour hopping}). Non-quadratic sources, if present, must be included in~(\ref{S_IV}). To better grasp the structure of $\mathbf{T}^{(V)}_{\mathrm{f/b}}(z)$ in~(\ref{contour hopping}), we emphasize that: 1) $\mathbf{T}_{\mathrm{f/b}}(t)$ is the physical time-dependent hopping matrix, so it depends on the real time coordinate $t$ and, if seen as a $z$-dependent quantity, it has the same value for $z = t_+$ and $z = t-$; 2) $\mathbf{V}_{\mathrm{f/b}}(z)$ is the source matrix, so it contributes only when $z$ is on the real-time branches and, for a given physical time $t$, it must satisfy $\mathbf{V}_{\mathrm{f/b}}(t_+) \neq \mathbf{V}_{\mathrm{f/b}}(t_-)$, so that $Z[V] \neq 1$; 3) $\mathbf{T}^{\mathrm{M}}_{\mathrm{f/b}}$ is constant on the Matsubara branch and does not contribute when $z$ is on the real-time branches, being replaced in this case by $\mathbf{T}_{\mathrm{f/b}}(t)$. 

Once this issue about the Matsubara Hamiltonian is taken into account, writing down the explicit form of equation~(\ref{S_IV}) is formally analogous to what is done in the case of the SK contour~\cite{Kamenev}. This term therefore presents no difficulties. In this Article, instead, we deal with the derivation of the explicit and unambiguous forms of~(\ref{S_f}) and~(\ref{S_b}). Indeed, the operator $i \partial_z $ appearing in these equations is 
a shorthand for an object that needs to be defined with great care within a discrete-time formulation. 
The difference between the KP $\gamma$ and SK $\gamma_{\rm K}$ time contours requires a generalization of the procedure given in~\cite{Kamenev}, which employs a discrete temporal mesh to properly take into account boundary conditions. Besides this, it should be kept in mind that, starting from the discrete temporal mesh, the path integral is well defined only in the limit of a vanishingly small time step, which is taken at the end of the derivation (see section~\ref{sect: discrete continuum}) and is not optional. With respect to~\cite{Kamenev}, in our derivation we also include an arbitrary matrix structure of the hopping matrix.

\section{The KP action}
\label{sect: KP action}

We now proceed to derive the path integral defining the nonequilibrium partition function on the KP time contour. We initially use the finite interval $[t_0, t_{\infty}]$ on the real-time axis (with $t_{\infty}>t_{0}$) and take the limit $t_{\infty} \rightarrow + \infty$ at the end of the derivation. 
This interval is divided into $N - 1$ arbitrarily small sub-intervals of width $\delta t$, so that $t_{\infty} - t_0 = (N - 1) \delta t$. We take the continuum $N \rightarrow \infty$ limit only at the end. We are therefore led to introduce a discrete set of time values:
\begin{eqnarray}
t_j \equiv t_0 + (j - 1) \delta t ~, \quad j = 1, 2, \ldots, N ~.
\label{physical time values}
\end{eqnarray}
Note that $t_1 = t_0$ and $t_N = t_{\infty}$. To each of the physical time values in~(\ref{physical time values}) we assign two distinct contour coordinates, $t_{j, +} \in \gamma_+$ and $t_{j, -} \in \gamma_-$.

We then consider the Matsubara branch $[t_0, t_0 - i \beta ]$. We split this interval into $M - 1$ infinitesimally small parts, such that $\beta = (M - 1) \delta t$, and we introduce
\begin{eqnarray}
\tau_j = t_0 - i (j - 1) \delta t ~, \quad j = 1, 2, \ldots, M ~.
\end{eqnarray}
We give a common name to the full set of discrete contour variables by introducing $2 N + M$ contour coordinates $z_j$ defined as
\begin{eqnarray}
z_j = t_{j, +} ~, \quad  j = 1, 2, \ldots, N ~, \nonumber \\
z_{N + j} = t_{( N + 1 - j), -} ~, \quad  j = 1, 2, \ldots, N ~, \nonumber \\
z_{ 2 N + j} = \tau_{j} ~, \quad  j = 1, 2, \ldots, M ~.
\end{eqnarray}

To determine the action $S^{(V)}[\overline{d}, d ; a^*, a]$ in equations~(\ref{ZKB}) and~(\ref{action KB}), which is a function of the fermionic (Grassmann) fields $(\bar{d}, d)$ and the bosonic (complex) fields $(a^*, a)$, we use a standard procedure that involves the decomposition of the identity operator of the full Hilbert space over fermionic and bosonic coherent states \cite{Kamenev, NegeleOrland, AltlandSimons}. Denoting by $a_j$, $a^*_j$, $\overline{d}_j$, $d_j$ the vectors collecting the fields needed to specify the Hamiltonian in the $j$-th decomposition of the identity, we write
\begin{eqnarray}
  \mathrm{Tr} \left\{ \hat{\mathcal{U}}^{(V)}_{\gamma}    \right\}    = 
\int \mathrm{d} \left[a^*_0, a_0 \right]   \int \mathrm{d} \left[\overline{d}_0, d_0 \right] \left< a_0, d_0 \right| \hat{\mathcal{U}}^{(V)}_{\gamma} \left| a_0, - d_0 \right> \mathrm{e}^{- \left| a_0 \right|^2}  \mathrm{e}^{- \overline{d}_0 \cdot d_0} \nonumber \\
  = 
\int \mathrm{d} \left[a^*_0, a_0 \right]   \int \mathrm{d} \left[\overline{d}_0, d_0 \right] \left< a_0, d_0 \right| \prod_{j = 1}^{2 N + M - 1} \hat{\mathcal{U}}^{(V)}_{z_j \rightarrow z_{j + 1}} \left| a_0, - d_0 \right> \mathrm{e}^{- \left| a_0 \right|^2}  \mathrm{e}^{- \overline{d}_0 \cdot d_0} \nonumber \\
 = 
\int \prod_{j = 1}^{2 N + M} \mathrm{d} \left[a^*_j, a_j \right]   \int \prod_{j = 1}^{2 N + M} \mathrm{d} \left[\overline{d}_j, d_j \right]  
\mathrm{e}^{- \sum_{j = 1}^{2 N + M} \left| a_j \right|^2}  \mathrm{e}^{- \sum_{j = 1}^{2 N + M} \overline{d}_j \cdot d_j} \nonumber \\
 \quad \times
        \int \mathrm{d} \left[a^*_0, a_0 \right]   \int \mathrm{d} \left[\overline{d}_0, d_0 \right]  \mathrm{e}^{- \left| a_0 \right|^2}  \mathrm{e}^{- \overline{d}_0 \cdot d_0}  \left< a_0, d_0  | a_{2 N + M}, d_{2 N + M} \right> \nonumber \\
        \quad \times \left( \prod_{j = 1}^{2 N + M - 1} \left< a_{j + 1}, d_{j + 1} \right| \hat{\mathcal{U}}^{(V)}_{z_j \rightarrow z_{j + 1}} \left| a_j, d_j \right> \! \right)   \left< a_1, d_1  | a_0, - d_0 \right>   \nonumber \\
 = 
\int \prod_{j = 1}^{2 N + M} \mathrm{d} \left[a^*_j, a_j \right]   \int \prod_{j = 1}^{2 N + M} \mathrm{d} \left[\overline{d}_j, d_j \right]  
\mathrm{e}^{- \sum_{j = 1}^{2 N + M} \left| a_j \right|^2}  \mathrm{e}^{- \sum_{j = 1}^{2 N + M} \overline{d}_j \cdot d_j} \nonumber \\
 \quad \times
        \mathrm{e}^{  a_1^* \cdot a_{2 N + M}}  \mathrm{e}^{- \overline{d}_1 \cdot d_{2 N + M}}   \left( \prod_{j = 1}^{2 N + M - 1} \left< a_{j + 1}, d_{j + 1} \right| \hat{\mathcal{U}}^{(V)}_{z_j \rightarrow z_{j + 1}} \left| a_j, d_j \right> \right)~.
\end{eqnarray}
Since $\left| a_1, d_1 \right>$ depends only on $a_1$ and $d_1$, and not on $a^*_1$ and $\overline{d}_1$, we notice that $\overline{d}_1$ enters the integral in the combination $\mathrm{e}^{- \overline{d}_1 \cdot \left( d_1 + d_{2 N + M} \right)}$ only, while $a^*_1$ enters the integral in the combination $\mathrm{e}^{- a^*_1 \cdot \left( a_1 - a_{2 N + M} \right)}$ only. Using the representations of ordinary and Grassmann $\delta$ functions~\cite{NegeleOrland}, we then observe that, for any function $f(a_1, d_1)$,
\begin{eqnarray}
\fl  \int \mathrm{d} \left[ a^*_1, a_1 \right]  \mathrm{e}^{- a^*_1 \cdot \left( a_1 - a_{2 N + M} \right)} \int \mathrm{d}[\overline{d}_1, d_1]  
\mathrm{e}^{- \overline{d}_1 \cdot ( d_1 + d_{2 N + M})}   f(a_1, d_1)     = f(a_{2N+M}, - d_{2 N + M} )~. 
\end{eqnarray}

We then obtain
\begin{eqnarray}
  \mathrm{Tr} \left\{ \hat{\mathcal{U}}^{(V)}_{\gamma}    \right\} 
& = 
\int \prod_{j = 2}^{2 N + M} \mathrm{d} \left[a^*_j, a_j \right]   \int \prod_{j = 2}^{2 N + M} \mathrm{d} \left[\overline{d}_j, d_j \right]   
\mathrm{e}^{- \sum_{j = 2}^{2 N + M}  \left( \left| a_j \right|^2 + \overline{d}_j \cdot d_j \right)} \nonumber \\
 &  \quad \times    \left( \prod_{j = 2}^{2 N + M - 1} \left< a_{j + 1}, d_{j + 1} \right| \hat{\mathcal{U}}^{(V)}_{z_j \rightarrow z_{j + 1}} \left| a_j, d_j \right> \right)   \nonumber \\
 &  \quad \times   \left< a_{2}, d_{2} \right| \hat{\mathcal{U}}^{(V)}_{z_1 \rightarrow z_2} \left| a_{2N+M}, - d_{2N+M} \right>   ~.
\end{eqnarray}
Taking $\delta t$ to be infinitesimally small, we have
\begin{eqnarray}
& \hat{\mathcal{U}}^{(V)}_{z_j \rightarrow z_{j + 1}} \rightarrow \hat{\mathcal{U}}^{(V)}_{+ \delta t} \approx \mathrm{e}^{- i \delta t \left[ \hat{\cal H}(t_j) + \hat{V}(t_{j, +}) \right] } , \quad \! 1 \leq j \leq N - 1 ~; \nonumber \\
& \hat{\mathcal{U}}^{(V)}_{z_N \rightarrow z_{N + 1}} = 1 ~; \nonumber \\
& \hat{\mathcal{U}}^{(V)}_{z_j \rightarrow z_{j + 1}} \rightarrow \hat{\mathcal{U}}^{(V)}_{- \delta t} \approx \mathrm{e}^{ i \delta t \left[ \hat{\cal H}(t_{2 N + 1 - j}) + \hat{V}(t_{2 N + 1 - j, -}) \right] } ~,   \quad \quad N + 1 \leq j \leq 2 N - 1 ~; \nonumber \\
& \hat{\mathcal{U}}^{(V)}_{z_{2 N} \rightarrow z_{2 N + 1}} = 1 ~; \nonumber \\
& \hat{\mathcal{U}}^{(V)}_{z_j \rightarrow z_{j + 1}} \rightarrow \hat{\mathcal{U}}^{(V)}_{- i \delta t} = \mathrm{e}^{ - \delta t \hat{\cal H}_{\rm M} } ~,  \quad \quad 2 N + 1 \leq j \leq     2 N + M - 1 ~.
\end{eqnarray}

We now introduce $\hat{\cal H}(z_j)$, $\hat{V}(z_j)$ and $\delta z_j$ in such a way that
\begin{eqnarray}
&   \hat{\cal H}(z_j) = \hat{\cal H}(t_j) ~, \quad \hat{V}(z_j) = \hat{V}(t_{j, +})  ~, \quad  1 \leq j \leq N - 1 ~; \nonumber \\
&   \hat{\cal H}(z_j) = \hat{\cal H}(t_{2 N + 1 - j}) ~, \quad \hat{V}(z_j) = \hat{V}(t_{2 N + 1 - j, -})  ~,  \quad\quad N+1 \leq j \leq 2 N - 1 ~; \nonumber \\
&  \hat{\cal H}(z_j) = \hat{\cal H}_{\rm M} ~, \quad \hat{V}(z_j) = 0 ~,  \quad\quad  2 N + 1 \leq j \leq 2 N + M - 1 ~;
\end{eqnarray}
and
\begin{eqnarray}
& \delta z_j = \delta t ~, \quad   1 \leq j \leq N - 1 ~; \nonumber \\
& \delta z_N = 0 ~;   \nonumber \\
& \delta z_j = - \delta t ~, \quad    N+1 \leq j \leq 2 N - 1 ~; \nonumber \\
& \delta z_{2 N} = 0 ~;   \nonumber \\
& \delta z_j = - i \delta t ~, \quad  2 N + 1 \leq j \leq 2 N + M - 1 ~.
\label{delta z}
\end{eqnarray}

We then write the contour evolution operator compactly as
\begin{eqnarray}
\hat{\mathcal{U}}^{(V)}_{z_j \rightarrow z_{j + 1}} \rightarrow \mathrm{e}^{- i \delta z_j \left[ \hat{\cal H}(z_j) + \hat{V}(z_j) \right] }~,
\end{eqnarray}
and we obtain
\begin{eqnarray}
\fl \mathrm{Tr} \left\{ \hat{\mathcal{U}}^{(V)}_{\gamma}    \right\}    = 
\int \prod_{j = 2}^{2 N + M} \mathrm{d} \left[a^*_j, a_j \right]   \int \prod_{j = 2}^{2 N + M} \mathrm{d} \left[\overline{d}_j, d_j \right]   \nonumber \\
  \times  \exp\left\{-  \sum_{j = 2}^{2 N + M  } \left[ a^*_{j } \cdot \left( a_{j  }  - a_{j - 1} \right)  + \overline{d}_{j} \cdot \left( d_{j} - d_{j - 1} \right) \right]              \right\}   \nonumber \\
  \times     \exp\left\{- i 
 \sum_{j = 1}^{2 N + M - 1} \delta z_j \left\{ {\cal H}\left[\overline{d}_{j + 1}, d_{j} ; a^*_{j + 1}, a_{j} ; z_j \right]   +    V\left[\overline{d}_{j + 1}, d_{j} ; z_j \right] \right\} \right\}~,
\label{path integral before continuum}
\end{eqnarray}
where $a_1 = a_{2N+M}$, and $d_1 = - d_{2N+M}$ in the arguments of the exponentials. In the continuous-time representation, one formally puts
\begin{eqnarray}
  \sum_{j = 2}^{2 N + M  } \left[ a^*_{j } \cdot \left( a_{j  }  - a_{j - 1} \right)  + \overline{d}_{j} \cdot \left( d_{j} - d_{j - 1} \right) \right]     \nonumber \\
    \equiv      \sum_{j = 2}^{2 N + M  }  \delta z_{j - 1} \left[ a^*(z) \cdot \partial_z a(z)     + \overline{d}(z) \cdot \partial_z   d(z) \right] \Big|_{z = z_j}~,
\end{eqnarray}
obtaining
\begin{eqnarray}
\mathrm{Tr} \left\{ \hat{\mathcal{U}}^{(V)}_{\gamma}    \right\}  \equiv  \int \mathcal{D} \left(\overline{d}, d \right) \int \mathcal{D} \left( a^*, a \right) \mathrm{e}^{i S^{(V)}\left[\overline{d}, d ; a^*, a \right]}~,
\label{path integral continuum}
\end{eqnarray}
where the action is given by equation~(\ref{action KB}). The boundary conditions $a_{1} = a_{2N + M}$ and $d_{1} = - d_{2 N + M}$ make the definition of the operator $\partial_z$ on the contour non-trivial, something which is fully captured by using the discrete temporal mesh. 

In passing, we mention that discrete-temporal-mesh methods have also been used to tackle other problems, related, for example, to the correct definition and use of time-derivative-dependent variable transformations in path integrals~\cite{Ezawa85}.

\section{Green's functions for the quadratic part of the action}
\label{sect: ff fb Gf}

\subsection{Derivation of the general formula}

In the discrete-time representation, the quadratic ($Q$) actions,~(\ref{S_f}) and~(\ref{S_b}), read as
\begin{eqnarray}
  S^{(V)}_{\mathrm{f}, Q}\!\left[  \overline{d} , d \right] =      \sum_{j, j' = 2}^{2 N + M} \overline{d}_{j}  \cdot \left( \mathbf{G}^{ -1}_{j, j' }   \right)^{(V)}_{\mathrm{f}, Q} d_{j'}     ~,
\label{free fermion discrete}
\end{eqnarray}
and
\begin{eqnarray}
  S^{(V)}_{\mathrm{b}, Q}\!\left[  a^*, a \right] =    \sum_{j, j' = 2}^{2 N + M} a^*_{j} \cdot \left( \mathbf{G}^{ -1}_{   j, j'  } \right)^{(V)}_{\mathrm{b} , Q} a_{ j'}     ~,
\label{free boson discrete}
\end{eqnarray}
where the free-fermion and free-boson inverse Green's functions (GFs) are matrices in the discrete-time coordinates ($j, j'$), as well as in the free-particle indexes. Note that we are using the term ``free'' in the sense of {\it non-interacting}. External fields coupling with the particles are included in the definition of $\mathbf{G}^{ -1}_Q$, but interactions are left into the action term $S_{\cal I}$, equation~(\ref{S_IV}). Our goal here is to find the direct GFs by inverting the matrices $\mathbf{G}^{ -1}_Q$. 

In order to lighten the notation, in the following steps we will omit subscripts and superscripts of the GFs, except for those referring to the discrete-time coordinates. To distinguish between bosons and fermions, we introduce the index $\xi = \pm 1$, where the $+$ sign is for bosons and the $-$ sign is for fermions. 

The free-particle inverse GF matrices are then written compactly as
\begin{eqnarray}
  i  \mathbf{G}^{-1}_{   j, j'  }        = -   \delta_{j, j'} \mathbf{1}    +  \left( 1 - \delta_{j, 2} \right) \delta_{j', j-1} \left( \mathbf{1} - i \delta z_{j-1}        \mathbf{ T}_{j - 1}   \right)      
  \nonumber \\
  \quad \quad\quad\,\,  + \xi  \delta_{j, 2}   \delta_{j', 2N+M }   \left( \mathbf{1} -  i \delta t  \mathbf{T}_{ 1}    \right) ~, 
     \label{Gfe-1} 
\end{eqnarray}
where the dependence of $\mathbf{T}_{j}$ on the contour coordinate is detailed in~(\ref{contour hopping}). The only formal difference between fermions and bosons is in the $\xi$-dependent term appearing in the second line of~(\ref{Gfe-1}), which reflects the different boundary conditions arising from the construction of the path integral.

The matrix in equation~(\ref{Gfe-1}) is invertible, i.e.~the left inverse is the same as the right inverse. We have
\begin{eqnarray}
  \sum_{j''} (   i   \mathbf{G}^{-1}_{j, j''}) ( - i \mathbf{G}_{j'', j'})   
  =  \delta_{j, j'} \nonumber \\
    \Rightarrow         i \mathbf{G}_{   j, j'  }       -  \left( 1 - \delta_{j, 2} \right)   \mathbf{x}_{j-1 }      i \mathbf{G}_{   j-1, j'  }     - \xi  \delta_{j, 2}      \mathbf{x}_{ 1 }   i \mathbf{G}_{   2N+M , j'  }        =  \delta_{j, j'} \mathbf{1} ~,
\label{Gfe definition} 
\end{eqnarray}
where
\begin{eqnarray}
\mathbf{x}_{j } \equiv \mathbf{1} - i \delta z_{j} \mathbf{T}_{j}~.
\end{eqnarray}
To solve equation~(\ref{Gfe definition}), we first introduce the Ansatz
\begin{eqnarray}
     i \mathbf{G}_{   j, j'  } =   \delta_{j, j'} \mathbf{1} -   \mathbf{X}_{   j, j'  }
    \label{ansatz}
\end{eqnarray}
in the second line of~(\ref{Gfe definition}); separating the case $j = 2$ from $j \neq 2$, we obtain
\begin{eqnarray}
         \mathbf{X}_{   2, j'  }           
        =  \xi     \mathbf{x}_{ 1 } \left( \mathbf{X}_{   2N+M , j'  }  - \delta_{j', 2N+M}  \mathbf{1}      \right) ~, \nonumber \\
        \mathbf{X}_{   j \neq 2, j'  }   
      =  \mathbf{x}_{j-1 }    \mathbf{X}_{   j-1, j'  }      - \delta_{j-1, j'}   \mathbf{x}_{j-1 }~.   
     \label{Gfe passage 2} 
\end{eqnarray}
The second equation ($j \neq 2$) is solved by iteration: for some $n < j-1$,
\begin{eqnarray}
  \mathbf{X}_{   j, j'  }   
         = \left( \overrightarrow{\prod_{m = 1}^{n} }  \mathbf{x}_{j-m   }   \right)    \mathbf{X}_{   j-n, j'  }   - \theta_{1 \leq j-j' \leq n}  \left( \overrightarrow{\prod_{m = 1}^{j-j'}} \mathbf{x}_{j-m }  \right) ,      \quad j >   n +  1~, 
     \label{Gfe passage 3} 
\end{eqnarray}  
where $\theta_p$ is the discrete step function ($\theta_p = 1$ if $p$ is true, $\theta_p = 0$ if $p$ is false), and we have introduced the contour-anti-ordered product
\begin{eqnarray}
\overrightarrow{\prod_{m = 1}^n} \mathbf{f}_{m} \equiv \mathbf{f}_1 \mathbf{f}_2 \ldots \mathbf{f}_{n-1} \mathbf{f}_n~,
\end{eqnarray}
which is accompanied by the contour-ordered product
\begin{eqnarray}
\overleftarrow{\prod_{m = 1}^n} \mathbf{f}_{m} \equiv \mathbf{f}_n \mathbf{f}_{n-1} \ldots \mathbf{f}_2 \mathbf{f}_1~.
\end{eqnarray}
From equation~(\ref{Gfe passage 3}), we take $n = j-2$ and, using the first among equations~(\ref{Gfe passage 2}), we find
\begin{eqnarray}
\fl   \mathbf{X}_{   j, j'  }   
        =   \xi   \left( \overleftarrow{\prod_{  m = 1 }^{    j-1}}   \mathbf{x}_{m  }   \right) 
        \left(      \mathbf{X}_{   2N+M , j'  }  - \delta_{j', 2N+M} \mathbf{1}     \right) 
          - \theta_{    j'\leq  j -1   }   \left( \overleftarrow{\prod_{m = j'}^{j - 1}}  \mathbf{x}_{m }  \right) ,     \quad j >   2~.
     \label{Gfe passage 4} 
\end{eqnarray}
We take $j = 2N+M$ and solve for every $j'$:
\begin{eqnarray}
    \mathbf{X}_{   2N+M , j'  }    = - \left(  \mathbf{1} - \xi   \mathbf{y}_{2N+M }   \right)^{-1}  \mathbf{y}_{2N+M }  \mathbf{y}^{-1}_{j' }
    + \delta_{j', 2N+M}  \mathbf{1}~, 
        \label{Xfe row 2N+M discrete} 
\end{eqnarray}
where we have introduced the quantities
\begin{eqnarray}
\mathbf{y}_{j } \equiv \left( \overleftarrow{\prod_{  m = 1 }^{j-1}  }   \mathbf{x}_{m  }   \right), \quad  
\mathbf{y}^{-1}_{j } = \left( \overrightarrow{\prod_{  m = 1 }^{j-1}  }   \mathbf{x}^{-1}_{m  }   \right) .
\end{eqnarray}
Inserting~(\ref{Xfe row 2N+M discrete}) into~(\ref{Gfe passage 4}), we solve the latter for $3 \leq j \leq 2N+M - 1$, while we solve for $j =2$ using~(\ref{Gfe passage 2}). The solutions for these cases can be combined into a single one:
\begin{eqnarray}
  \mathbf{X}_{   j, j'  }   
           =         \mathbf{y}_{j }     \left[ - \xi   \left(  \mathbf{1} - \xi   \mathbf{y}_{2N+M }   \right)^{-1}    \mathbf{y}_{2N+M }               - \theta_{    j'\leq  j -1   } \mathbf{1}  \right]   \mathbf{y}^{-1}_{j' }  .
     \label{Xfe discrete} 
\end{eqnarray}
It can be seen that~(\ref{Xfe discrete}) coincides with~(\ref{Xfe row 2N+M discrete}) for $j = 2N+M$. Therefore,~(\ref{Xfe discrete}) represents the full solution for all values of $j, j'$. From~(\ref{ansatz}), we find the free-particle GF:
\begin{eqnarray}
     i \mathbf{G}_{   j, j'  } =       \mathbf{y}_{j }    \left[  \xi   \left( \mathbf{y}_{2N+M }^{-1}  - \xi \mathbf{1}     \right)^{-1}        + \theta_{    j'\leq  j    }  \mathbf{1} \right]  \mathbf{y}^{-1}_{j' } ~. 
    \label{G discrete}
\end{eqnarray}
We now take the continuum limit. We first introduce
\begin{eqnarray} \label{times}
& t_j = t_0 + (j - 1) \delta t~, \quad \mathrm{if} \,\, 1 \leq j \leq N ~; \nonumber \\
& t_j = t_0 + (2 N - j) \delta t~, \quad \mathrm{if} \,\,   N + 1 \leq j \leq 2N ~; \nonumber \\
& t_j = t_0 - i (j - 2 N - 1) \delta t~,  \quad \mathrm{if} \,\,   2 N + 1 \leq j \leq 2N + M~.
\end{eqnarray}
These identities give the real (if $1 \leq j \leq 2N$) or complex (if $2 N + 1 \leq j \leq 2N + M$) time coordinates on the contour corresponding to the discrete index $j$.

Taking $\delta t$ as infinitesimally small, we find
\begin{eqnarray} \label{yj continuum 1}
\mathbf{y}_{j }   \rightarrow \mathcal{T}_{\gamma} \exp \left[ - i  \int_{t_{0+}}^{z_j} \mathrm{d} z' \, \mathbf{T}(z') \right]
\end{eqnarray}
and
\begin{eqnarray} \label{yj continuum 2}
\mathbf{y}_{j }^{-1}   \rightarrow \overline{\mathcal{T}_{\gamma}} \exp \left[  i  \int_{t_{0+}}^{z_j} \mathrm{d} z' \, \mathbf{T}(z') \right]~,   
\end{eqnarray}
where $\mathcal{T}_{\gamma}$ is the contour-ordering operator along $\gamma$, and $\overline{\mathcal{T}_{\gamma}}$ is the analogous contour-anti-ordering operator. In particular,
\begin{eqnarray} \label{towards the number matrix}
   \mathbf{y}_{2 N + M }^{-1}    \rightarrow \overline{\mathcal{T}_{\gamma}} \exp \left[  i  \int_{\gamma} \mathrm{d} z' \, \mathbf{T}(z') \right]  \nonumber \\
      =  \left\{ \overline{\mathcal{T}_{\gamma}} \exp \left[  i  \int_{t_{0+}}^{t_{0-}} \mathrm{d} z' \, \mathbf{T}(z') \right]  \right\} \exp\left( \beta \mathbf{T}^{\mathrm{M}}\right)~,   
\end{eqnarray}
where the simplification is possible due to the contour structure of the hopping detailed in equation~(\ref{contour hopping}). In the second step of~(\ref{towards the number matrix}), the contour-anti-ordered exponential is different from $\mathbf{1}$ only in the presence of sources, while the term $\exp\left( \beta \mathbf{T}^{\mathrm{M}}\right)$ is a matrix generalization of the inverse Boltzmann factor. This suggests to introduce a generalized occupation-number matrix,
\begin{eqnarray} \label{number matrix with sources}
\mathbf{n}_{\xi}    \equiv \left( \mathbf{y}_{2N+M }^{-1}  - \xi \mathbf{1}     \right)^{-1}    \nonumber \\
\quad \,\,  =      \left\{ \overline{\mathcal{T}_{\gamma}} \exp \left[  i  \int_{t_{0+}}^{t_{0-}} \mathrm{d} z' \, \mathbf{T}(z') \right]   \exp\left( \beta \mathbf{T}^{\mathrm{M}}\right)  - \xi \mathbf{1}     \right\}^{-1}~.
\end{eqnarray}
We will see that, in the simplest cases, this quantity reduces to the standard occupation number.

We now switch to the full continuum notation for the GF, obtaining
\begin{eqnarray}
     i \mathbf{G}_{\xi}(z, z') & =        \left\{ \mathcal{T}_{\gamma} \exp\left[ - i \int_{t_{0+}}^{z} \mathrm{d} z'' \, \mathbf{T}(z'')   \right]     \right\}    \left[ \xi  \mathbf{n}_{\xi}         + \Theta(z, z') \mathbf{1} \right] \nonumber \\
 & \quad \times  \left\{ \overline{\mathcal{T}_{\gamma}} \exp\left[ i \int_{t_{0+}}^{z'} \mathrm{d} z'' \, \mathbf{T}(z'')   \right]     \right\} ~,
    \label{G discrete continuum limit}
\end{eqnarray}
where $\Theta(z,z')$ is the step function on $\gamma$, with $\Theta(z,z) = 1$. 

Equation~(\ref{G discrete continuum limit}) represents the explicit nonequilibrium free-fermion and free-boson GFs on the KP time contour $\gamma$, for the most general case of a single-particle Hamiltonian with a non-trivial matrix structure, including time-dependent fields {\it and} sources. We note that, in the absence of sources ($V = 0$), equation~(\ref{G discrete continuum limit}) reduces to the noninteracting GF derived in~\cite{Stefanucci} with the method of equations of motion. We will now consider several relevant cases in which~(\ref{G discrete continuum limit}) can be significantly simplified. 

\subsection{GFs in the absence of sources}
\label{subsect: no V}

Most nonequilibrium problems require functional differentiation with respect to the sources, followed by the evaluation of the result at $\mathbf{V}(z) = \mathbf{0}$. Therefore, although nonzero sources are needed to make the path integral meaningful, at some point in the calculation one typically needs to compute the GFs in the absence of sources. In this case, $\mathbf{T}(z) = \mathbf{T}(t)$ if $z \in \gamma_+ \cup \gamma_-$, while $\mathbf{T}(z) = \mathbf{T}^{\mathrm{M}}$ if $z \in \gamma_{\mathrm{M}}$. Equations~(\ref{yj continuum 1}) and (\ref{yj continuum 2}) simplify to
\begin{eqnarray} \label{simplify 1 1}
\mathbf{y}_{j }   = \mathcal{T} \exp \left[ - i  \int_{t_0}^{t_j} \mathrm{d} t' \, \mathbf{T}(t') \right] \quad \mathrm{if} \, z_j \in \gamma_+ \cup \gamma_- ~,
\end{eqnarray}
\begin{eqnarray} \label{simplify 1 2}
\mathbf{y}_{j }   =   \exp \left( - \tau \mathbf{T}^{\mathrm{M}} \right) \quad \mathrm{if} \, z_j \in \gamma_{\mathrm{M}} ~,
\end{eqnarray}
\begin{eqnarray} \label{simplify 2 1}
\mathbf{y}_{j }^{-1}   = \overline{\mathcal{T}} \exp \left[ i  \int_{t_0}^{t_j} \mathrm{d} t' \, \mathbf{T}(t') \right] \quad \mathrm{if} \, z_j \in \gamma_+ \cup \gamma_- ~,
\end{eqnarray}
and
\begin{eqnarray}\label{simplify 2 2}
\mathbf{y}_{j }^{-1}   =   \exp \left(  \tau \mathbf{T}^{\mathrm{M}} \right) \quad \mathrm{if} \, z_j \in \gamma_{\mathrm{M}} ~, 
\end{eqnarray}
where $\mathcal{T}$ and $\overline{\mathcal{T}}$ are the standard time-ordering and anti-time-ordering operators on the real axis, respectively, $t_j$ are the time coordinates defined as in~(\ref{times}), and $0< \tau < \beta$ parameterizes the Matsubara branch as $t_0 - i \tau$. In particular,
\begin{eqnarray}
  \mathbf{y}_{ 2 N + M }   =   \exp \left( - \beta \mathbf{T}^{\mathrm{M}} \right)~. 
\end{eqnarray}
The occupation number matrix introduced in~(\ref{number matrix with sources}) becomes 
\begin{eqnarray}
\mathbf{n}_{\xi} \equiv \left(   \mathrm{e}^{  \beta   \mathbf{T}^{\mathrm{M}} }      -  \xi \mathbf{1} \right)^{-1}~,
\label{occupation number matrix}
\end{eqnarray}
which is the matrix generalization of the occupation number resulting from either the Bose-Einstein or the Fermi-Dirac statistics. Once $z$ and $z'$ are specified, equation~(\ref{G discrete continuum limit}) is simplified using~(\ref{simplify 1 1})-(\ref{simplify 2 2}) and~(\ref{occupation number matrix}).

\subsection{GFs in the equilibrium case for a diagonal single-particle Hamiltonian}

Let us consider the case of an equilibrium single-particle GF and assume that the single-particle Hamiltonian (and, therefore, the GF as well) is diagonal in the single-particle indexes. It should be noted that treating an equilibrium single-particle GF does not mean that the system must be at equilibrium. It only means that one chooses to exclude the nonequilibrium features from the definition of the single-particle Hamiltonian, i.e.~the time-dependent fields are included into~(\ref{S_IV}), despite being quadratic contributions to the action. Let $k$ be the set of single-particle quantum numbers. For a system of electrons on a lattice, such set would consist of the wavevector and a spin projection.

Including a possibly $k$-dependent chemical potential term (e.g.~a spin-dependent chemical potential for electrons), which enters the definition of the Hamiltonian on the Matsubara branch, we find $T_{k , k'}(z) = \delta_{k, k'}   \varepsilon_{k }(z)$, with
\begin{eqnarray}
\varepsilon_{ k }(z) = \varepsilon_{ k } - \mu_{k} \Theta(z, t_{0-})~,
\end{eqnarray}
and the number matrix reduces to the familiar occupation number,
\begin{eqnarray}
n_{\xi; k  } =   \left(   \mathrm{e}^{  \beta \left( \varepsilon_{ k }   - \mu_{k} \right)}      -  \xi \right)^{-1} ~.
\end{eqnarray}

In this case, the free-particle GF is diagonal, with diagonal components given by
\begin{eqnarray}
G_{\xi ; k } (z,z') = & - i \, \mathrm{e}^{- i  \left[ \varepsilon_{ k }(z) \left( t  - t_0 \right)    -  \varepsilon_{ k }(z') \left( t'  - t_0 \right) \right] }      \left[    \xi \, n_{\xi ; k}   +  \Theta(z,z')    \right] ~,
\label{Gf continuum}
\end{eqnarray}
where $t$ and $t'$ are the (complex) time coordinates corresponding to the contour coordinates $z$ and $z'$, respectively. Recall that $\xi=\pm 1$ distinguishes between the bosonic and fermionic cases, respectively. Equation~(\ref{Gf continuum}) represents the generalization of the results given in~\cite{Kamenev} for the SK time contour, to the case of the KP time contour.

To better illustrate the main features of equation~(\ref{Gf continuum}), we explicitly consider the various combinations obtained when the positions of $z$ and $z'$ on the contour are specified. First, consider the case in which both $z$ and $z'$ belong to the real-time branches (either to $\gamma_+$ or to $\gamma_-$). In this case, one obtains
\begin{eqnarray}
G^{  >}_{\xi; k } (t,t') =   - i \,  \mathrm{e}^{- i    \varepsilon_{ k }  \left( t  - t' \right)      }   \left( 1   + \xi n_{\xi ; k}      \right)~,
\end{eqnarray}
if $z > z'$ (on the contour) and
\begin{eqnarray}
G^{<}_{\xi ; k } (t,t') =  - \xi \,   i \,  \mathrm{e}^{- i    \varepsilon_{ k}  \left( t  - t' \right)      }     n_{\xi; k}~,
\end{eqnarray}
if $z < z'$. These coincide with the standard {\it greater} and {\it lesser} nonequilibrium GFs which are also found in the SK formalism~\cite{Kamenev}.

On the KP contour, one also obtains the other Langreth components~\cite{Stefanucci}. If $z$ belongs to one of the two real-time branches, while $z'$ is on the Matsubara branch,
\begin{eqnarray}
G^{ \rceil}_{\xi ; k} (t,t_0 - i \tau) =  - \xi \,   i \,  \mathrm{e}^{- i    \varepsilon_{ k }  \left( t  - t_0 \right)   }   \mathrm{e}^{ \left(   \varepsilon_{ k }   - \mu_k \right) \tau   }   n_{\xi ; k}~.
\end{eqnarray}
In the opposite case, when $z$ belongs to the Matsubara branch and $z'$ to one of the real-time branches,
\begin{eqnarray}
G^{  \lceil}_{\xi ; k } (t_0 - i \tau, t) & =     - i \,  \mathrm{e}^{  i    \varepsilon_{ k }  \left( t  - t_0 \right)   }   \mathrm{e}^{ - \left(   \varepsilon_{ k }    - \mu_k \right) \tau   }   \left( 1 + \xi  n_{\xi ; k}   \right) ~.
\end{eqnarray}
Finally, if both $z$ and $z'$ belong to the Matsubara branch,
\begin{eqnarray}
G^{ \rm M}_{\xi ; k } (t_0 - i \tau, t_0 - i \tau') & =     - i \,  \mathrm{e}^{  - \left(    \varepsilon_{k}   - \mu_k \right) \left( \tau  - \tau' \right)   }      \left[ \theta(\tau - \tau') + \xi  n_{\xi ; k}   \right]~,
\end{eqnarray}
where $\theta(\tau- \tau')$ is the ordinary step function.

\subsection{Determinant}
\label{sect: determinant}

In several applications which involve field integration over either the fermionic/bosonic degrees of freedom, it is necessary to know the determinant of $( -  i  \mathbf{G}^{-1}_{   j, j'  })$. In general, explicit calculations of this quantity are difficult. 

However, in the case when the hopping is diagonal, the calculation can be done easily and directly from equation~(\ref{Gfe-1}). The determinant can be evaluated using the Laplace theorem combined with the fact that the determinant of a triangular matrix is the product of its diagonal elements. The result is independent of $M$ and $N$ being even or odd, implying that it is well defined in the limit $M, N \rightarrow \infty$. We obtain (omitting the single-particle quantum numbers)
\begin{eqnarray}
  \mathrm{det}( -  i  G^{-1}_{   j, j'  })  = 1 - \xi \prod_{j = 1}^{2 N + M - 1} x_j =  1 - \xi y_{2 N + M }     \rightarrow  1 -  \xi \mathrm{e}^{- \beta \left( \varepsilon - \mu \right) } ~.
\label{determinant A}
\end{eqnarray}

\section{From the discrete to the continuum representation}
\label{sect: discrete continuum}

We now apply the continuum representation for the time domain, in order to make the path-integral construction meaningful. We define the operator
\begin{eqnarray}
\hat{\mathbf{G}}^{ -1}_{\xi}(z,z')  \equiv 
\frac{\left( \mathbf{G}^{  -1}_{ \xi}\right)_{   j, j'  }}{\delta z_{j-1} \delta z_{j'-1}}~,
\label{Gfe-1 continuum}
\end{eqnarray}
which satisfies
\begin{eqnarray}
   \int_{\gamma} \mathrm{d}z' \, \hat{\mathbf{G}}^{-1}_{\xi}(z,z')~\mathbf{G}_{\xi}(z',z'')   = \mathbf{1} \delta(z, z'')~,
\label{Gfe-1 continuum definition}
\end{eqnarray}
where $\mathbf{G}_{\xi}(z',z'')$ is given by~(\ref{G discrete continuum limit}) and $\delta(z, z'')$ is the Dirac delta on $\gamma$, namely
\begin{eqnarray}
\delta(z, z'') \equiv \frac{\delta_{j, j''}}{\delta z_{j-1}}~.
\end{eqnarray}
In the continuum, equation~(\ref{free fermion discrete}) reads as   
\begin{eqnarray}
  S_{\mathrm{f}, Q}\left[  \overline{d} , d \right] 
  = 
        \int_{\gamma}   \mathrm{d}z  \int_{\gamma}  \mathrm{d}z' \,      \overline{d}(z) \cdot \hat{\mathbf{G}}^{ -1 }_{\mathrm{f}, Q}(z,z')  \,        d(z')~,
\label{fermions continuum}
\end{eqnarray}
while equation~(\ref{free boson discrete}) reads as
\begin{eqnarray}
  S_{\mathrm{b}, Q}\left[  a^* , a \right] 
  = 
       \int_{\gamma}   \mathrm{d}z   \int_{\gamma} \mathrm{d}z'     a^*(z) \cdot \hat{\mathbf{G}}^{  -1  }_{\mathrm{b}, Q}(z,z')   \,  a(z')~.
           \label{bosons continuum}
\end{eqnarray}
The dependence on the sources $V$ is implicit. The inverse GFs appearing in~(\ref{fermions continuum}) and~(\ref{bosons continuum}) are to be considered as merely symbolic objects. Any calculation (involving, for example, field integration or extremization of the action) must, at some stage, rely on a transformation from the inverse GF to the direct GF in equation~(\ref{G discrete continuum limit}). This encodes the boundary conditions arising from the procedure of construction of the path integral, while being fully well-defined in the continuum limit of the time domain.

\section{An example: Spin evolution in a time-dependent Zeeman field}
\label{sect:example}

To show how the KP time contour can be used to specify the initial state in practice, we consider a simple and well-known problem that can be solved analytically with elementary means, allowing for a straightforward comparison of the result with that obtained in the realm of path integrals.

The model involves a single electronic orbital subjected to a time-dependent magnetic field along a certain direction $x$, which couples to the spin through the Zeeman coupling. 
We neglect interactions, which would affect the doubly-occupied state and choose a basis for the electronic states where spin is quantized along $z$, with $z \perp x$. The Hamiltonian is
\begin{eqnarray}
\hat{\cal H}(t) \equiv   B_x(t) \hat{s}_x~,
\end{eqnarray}
where $\hat{s}_x = ( \hat{d}^{\dagger}_{\uparrow} \hat{d}_{\downarrow} + \hat{d}^{\dagger}_{\downarrow} \hat{d}_{\uparrow}) / 2$ and $B_x(t)$ is the external magnetic field, in appropriate units. We assume $B_x(t_0) = 0$. Our goal is to compute the time-dependent density matrix
\begin{eqnarray}
\rho_{\sigma, \sigma'}(t) \equiv \left\langle \hat{\cal U}(t_0, t) \, \hat{d}^{\dagger}_{\sigma} \hat{d}_{\sigma'} \, \hat{\cal U}(t, t_0) \right\rangle~, 
\label{ex rho}
\end{eqnarray}
from which single-particle observables (particle number and spin components) can be read off. The quantity~(\ref{ex rho}) is obviously very sensitive to the initial preparation of the system, demonstrating spin precession if the spin in the initial state has a non-vanishing component perpendicular to $x$, or no evolution at all otherwise. In section~\ref{subsect: QM} we compute~(\ref{ex rho}) by using tools of basic quantum mechanics. Then, in section~\ref{subsect: path} we compute~(\ref{ex rho}) by using the KP partition function that we have derived before, to show that an appropriate choice of $\hat{\cal H}_{\mathrm M}$ allows to choose the initial state and therefore to reproduce all possible dynamical evolutions of the system.

\subsection{Solution of the problem with ordinary quantum mechanics}
\label{subsect: QM}

We first solve the problem using ordinary quantum mechanics. The Fock space consists of four states: the vacuum $\left| 0 \right>$, the singly-occupied states $\hat{d}^{\dagger}_{\sigma} \left| 0 \right> \equiv \left| \sigma \right>$, and the doubly-occupied state $\hat{d}^{\dagger}_{\uparrow} \hat{d}^{\dagger}_{\downarrow} \left| 0 \right> \equiv \left| \uparrow \downarrow \right>$. The evolution operator, satisfying $i \, \partial_t \, \hat{\cal U}(t, t_0)  = \hat{\cal H}(t) \, \hat{\cal U}(t, t_0) $ and $\hat{\cal U}(t_0, t_0) = \hat{1}$, is obtained directly as
\begin{eqnarray}
\hat{\cal U}(t, t_0) & = \left| 0 \right>   \left< 0 \right|  
+  \left| \uparrow \downarrow \right>   \left< \uparrow \downarrow \right|   + \left\{ \cos\left[\theta(t) / 2\right] \left| \uparrow \right>  - i \sin\left[\theta(t) / 2\right] \left| \downarrow \right> \right\} \left< \uparrow \right| \nonumber \\
& \quad + \left\{ \cos\left[\theta(t) / 2\right] \left| \downarrow \right>  - i \sin\left[\theta(t) / 2 \right] \left| \uparrow \right> \right\} \left< \downarrow \right|~,
\label{ex U}
\end{eqnarray}
where
\begin{eqnarray}
\theta(t) \equiv   \int^t_{t_0} \mathrm{d} t' B_x(t')~.
\label{ex theta}
\end{eqnarray}
The expectation value in~(\ref{ex rho}) can be calculated from
\begin{eqnarray}
\rho_{\sigma, \sigma'}(t) \equiv \sum_{n} W_n \left\langle n \right| \hat{\cal U}(t_0, t) \, \hat{d}^{\dagger}_{\sigma} \hat{d}_{\sigma'} \, \hat{\cal U}(t, t_0) \left| n \right\rangle~,
\end{eqnarray}
where $W_n$ is the statistical weight of the Fock state $\left| n \right>$. The final result (after a number of simple trigonometric manipulations) is
\begin{eqnarray}
\rho_{\uparrow, \uparrow}  =  W_{\uparrow \downarrow}      +     \frac{1}{2} \left( W_{\uparrow} + W_{\downarrow} \right)  + \frac{1}{2} \left( W_{\uparrow} - W_{\downarrow} \right)  \cos\left[  \theta(t)  \right]~, \nonumber \\
\rho_{\uparrow, \downarrow} =  - \frac{i}{2}  \left( W_{\uparrow}  - W_{\downarrow}  \right)  \sin\left[   \theta(t) \right]~, \nonumber \\
\rho_{\downarrow, \uparrow} =  \frac{i}{2} \left( W_{\uparrow}  - W_{\downarrow}  \right)  \sin\left[   \theta(t) \right]~, \nonumber \\
\rho_{\downarrow, \downarrow}  =  W_{\uparrow \downarrow}     +    \frac{1}{2} \left( W_{\uparrow} + W_{\downarrow} \right) - \frac{1}{2} \left( W_{\uparrow} - W_{\downarrow} \right) \cos\left[  \theta(t)  \right]~.
\label{ex rho QM}
\end{eqnarray}
It is important to notice that the freedom on the choice of the weights $W_n$ allows to prepare the system in a non-trivial initial state and to implement symmetry breaking. For example, if we make the rotationally symmetric choice $W_{\uparrow} = W_{\downarrow}$, we get $\rho_{\sigma, \sigma'}(t) = \delta_{\sigma, \sigma'} \left( W_{\uparrow \downarrow} + W_{\uparrow} \right)$: the density matrix becomes time-independent, encoding only information about the (constant) population of the orbital, as all the spin components are zero. This choice, in more sophisticated cases, is not uncommon at all: it is exactly what one would obtain from a grand canonical thermal mixture, where $W_n = \mathrm{e}^{- \beta (E_n - \mu N_n)} / \mathcal{Z}$ and $\mathcal{Z} = \sum_n \mathrm{e}^{- \beta (E_n - \mu N_n)}$. In such case, the equality between $W_{\uparrow}$ and $W_{\downarrow}$ comes from the degeneracy of the states $\left| \uparrow \right>$ and $\left| \downarrow \right>$ with respect to the Hamiltonian $\hat{\cal H}(t_0)$. It is by allowing $W_{\uparrow} \neq W_{\downarrow}$ (hence, deviating from a thermal mixture) that rotational symmetry is broken, the $z$ direction is selected as the one along which the spin is aligned in the initial state, and a non-trivial spin evolution occurs in response to the application of a magnetic field perpendicular to $z$.

\subsection{Solution of the problem by means of the KP path integral approach}
\label{subsect: path}

Can we obtain the same result (\ref{ex rho QM}) within the KP path integral formulation? In particular, do we have the same freedom on the preparation of the initial state? To see this, let us follow the procedure outlined above. Recalling that $B_x(t_0) = 0$, we take the following Hamiltonian on the Matsubara branch:
\begin{eqnarray}\label{Matsubara spin}
\hat{\cal H}_{\rm M} =   - \mu \left( \hat{n}_{\uparrow} + \hat{n}_{\downarrow} \right) + \sum_{\sigma} (\sigma \Delta)~\hat{n}_{\sigma}~,
\end{eqnarray}
where $\hat{n}_{\sigma}=\hat{d}^\dagger_{\sigma} \hat{d}_{\sigma}$. The inclusion of the second term ($\propto \Delta$) in equation~(\ref{Matsubara spin}), which is equivalent to assuming a spin-dependent chemical potential, is the key tool that allows to break rotational symmetry in the initial-state preparation. We emphasize that~(\ref{Matsubara spin}), for $\Delta \neq 0$, is crucially different from the grand-canonical Matsubara Hamiltonian, which would be equal to $- \mu \left( \hat{n}_{\uparrow} + \hat{n}_{\downarrow} \right)$. Therefore, we are {\it not} choosing a thermal mixture. 

The nonequilibrium partition function, from equation~(\ref{ZKB}), is
\begin{eqnarray}
Z\left[ V \right] =  \mathcal{Z}^{-1}     \int \mathcal{D} \left(\overline{d}, d \right)   \mathrm{e}^{i S^{(V)}\left[\overline{d}, d  \right]}~,
\label{ex ZKB}
\end{eqnarray}
where 
\begin{eqnarray}
\mathcal{Z}  = \mathrm{Tr}\!\left( \mathrm{e}^{- \beta \hat{\cal H}_{\mathrm{M}}}\right) =  1 + \mathrm{e}^{\beta (\mu - \Delta)}  + \mathrm{e}^{\beta (\mu + \Delta)} + \mathrm{e}^{2 \beta   \mu } 
\label{Z equilibrium spin problem}
\end{eqnarray}
is the equilibrium partition function. Because $Z[0] = 1$, we have
\begin{eqnarray}
\int \mathcal{D} \left(\overline{d}, d \right)   \mathrm{e}^{i S^{(0)}\left[\overline{d}, d  \right]} =  \mathcal{Z}~.
\end{eqnarray}

The action is quadratic, as in~(\ref{fermions continuum}). The free-electron GF has the general form given by~(\ref{G discrete continuum limit}), with $\mathbf{T}$ being a matrix in spin space only:
\begin{eqnarray}
\mathbf{T}(z) & = \left( \bm{\sigma}_3 \Delta - \mathbf{1} \mu \right)  \Theta(z, t_{0-})  + \bm{\sigma}_1 \frac{B_x(t)}{2}  \Theta(t_{0-}, z) \nonumber \\
& \equiv   \mathbf{T}^{\mathrm{M}} \, \Theta(z, t_{0-})  + \mathbf{T}(t) \,  \Theta(t_{0-}, z)~,
\end{eqnarray}
where $\mathbf{1}$ is the $2 \times 2$ identity and $\bm{\sigma}_{n}$ are ordinary $2\times2$ Pauli matrices. The source term in the action is
\begin{eqnarray}
V\left[ \overline{d}(z), d(z)  ; z  \right] \equiv \sum_{\sigma, \sigma' }  V_{\sigma, \sigma'}(z) \, \overline{d}_{\sigma}(z) \, d_{\sigma'}(z)~,
\end{eqnarray}
so that the density matrix is obtained from functional differentiation with respect to the source fields as
\begin{eqnarray}
\rho_{\sigma, \sigma'}(t) = \frac{i}{2} \left. \left\{ \frac{\delta Z[V]}{\delta V_{\sigma, \sigma'}(t_+) }  
+  \frac{\delta Z[V]}{\delta V_{\sigma, \sigma'}(t_-) } \right\} \right|_{V = 0}~.
\label{ex rho path integral}
\end{eqnarray}

Carrying out the path integral in~(\ref{ex ZKB}), we obtain
\begin{eqnarray}
  Z_0\left[ V \right]  = \mathcal{Z}^{-1}    \mathrm{det} \left( - i \, G^{-1}[V] \right)  
  = \mathcal{Z}^{-1}   \mathrm{e}^{ \mathrm{tr} \, \mathrm{ln} \left( - i \, G^{-1}[V] \right) }  ~.
\label{ex ZKB - 0}
\end{eqnarray}
The explicit calculation of~(\ref{ex rho path integral}) requires the following identity:
\begin{eqnarray}
 \fl \mathrm{tr}  \left.   \frac{\delta  \,  \mathrm{ln} \left( - i \, G^{-1}[V] \right)  }{\delta V_{\sigma, \sigma'}(z) }  
     \right|_{V = 0}  
       = \mathrm{tr}\left\{  G[0]  \left.   \frac{\delta     \left(   G^{-1}[V] \right)  }{\delta V_{\sigma, \sigma'}(z) }  
     \right|_{V = 0} \right\}    = - G_{\sigma', \sigma}(z, z + 0) ~,
\end{eqnarray}
where $V = 0$ is intended in the last line. We finally find
\begin{eqnarray}
\rho_{\sigma, \sigma'}(t) = -  i  G_{\sigma', \sigma}[t_+, (t + 0)_+] \equiv  -  i  G^<_{\sigma', \sigma}(t, t)  ~.
  \label{ex rho path integral nonint}
\end{eqnarray}

To calculate the GF, let us consider equation~(\ref{G discrete continuum limit}) specialized to our case. Since the GF needs to be calculated in the absence of sources, we can use the simplifications discussed in section~\ref{subsect: no V}. We find
\begin{eqnarray}
  -   i \mathbf{G}^<(t, t) & =        \left\{ \mathcal{T}  \exp\left[ - i \int_{t_0}^{t} \mathrm{d} t' \, \mathbf{T}(t')   \right]     \right\}     \mathbf{n}_{\mathrm{f}}        \left\{ \overline{\mathcal{T} } \exp\left[ i \int_{t_0}^{t} \mathrm{d} t' \, \mathbf{T}(t')   \right]     \right\}  \nonumber \\
 & =             \exp\left[ - i  \bm{\sigma}_1    \theta(t) / 2  \right]     \mathbf{n}_{\mathrm{f}}      \exp\left[   i  \bm{\sigma}_1      \theta(t)  / 2 \right]  ~,
    \label{ex G}
\end{eqnarray}
where we have exploited the fact that $\mathbf{T}(t)$ commutes with itself at different times.  The exponentials of the Pauli matrix give the usual result:
\begin{eqnarray}
\exp\left[ \pm  i  \bm{\sigma}_1   \theta(t) / 2 \right] = \mathbf{1} \cos\left[     \theta(t) / 2     \right] \pm i \bm{\sigma}_1 \sin\left[     \theta(t) / 2   \right]~.
\end{eqnarray}
Equation~(\ref{ex G}) includes information about the preparation of the initial state via the occupation number matrix:
\begin{eqnarray} 
\left( \mathbf{n}_{\mathrm{f}} \right)_{\sigma, \sigma'} = \left[ \exp\left( \beta \mathbf{T}^{\mathrm{M}} \right) + \mathbf{1} \right]^{-1}_{\sigma, \sigma'} = \delta_{\sigma, \sigma'}   \left(  \mathrm{e}^{\beta \left( \sigma\Delta - \mu \right)} + 1 \right)^{-1} \equiv \delta_{\sigma, \sigma'} \,  n_{\mathrm{f}, \sigma}~.
\label{occupation matrix}
\end{eqnarray}
The components of~(\ref{ex G}), after simple trigonometric manipulations, can be written as
\begin{eqnarray}
  -   i G_{\uparrow, \uparrow}^<(t, t)  =      \frac{1}{2} \left( n_{\mathrm{f}, \uparrow}  + n_{\mathrm{f}, \downarrow} \right) +  \frac{1}{2} \left( n_{\mathrm{f}, \uparrow}  - n_{\mathrm{f}, \downarrow} \right) \cos\left[ \theta(t)   \right]\equiv \rho_{\uparrow, \uparrow}(t)~, \nonumber \\
  -   i G_{\uparrow, \downarrow}^<(t, t) = \frac{i}{2} \left( n_{\mathrm{f}, \uparrow}  - n_{\mathrm{f}, \downarrow} \right)  \sin\left[   \theta(t) \right]  \equiv \rho_{\downarrow, \uparrow}(t)~,  \nonumber \\
  -   i G_{\downarrow, \uparrow}^<(t, t) = - \frac{i}{2} \left(n_{\mathrm{f}, \uparrow}   -     n_{\mathrm{f}, \downarrow}  \right)  \sin\left[   \theta(t) \right]  \equiv \rho_{\uparrow, \downarrow}(t)~, \nonumber \\
  -   i G_{\downarrow, \downarrow}^<(t, t)  
=   \frac{1}{2} \left( n_{\mathrm{f}, \uparrow} + n_{\mathrm{f}, \downarrow} \right)  - \frac{1}{2} \left( n_{\mathrm{f}, \uparrow} - n_{\mathrm{f}, \downarrow} \right) \cos\left[  \theta(t)  \right]   
        \equiv \rho_{\downarrow, \downarrow}(t)~,
    \label{ex G final}
\end{eqnarray}
where we have identified the components of the density matrix $\rho_{\sigma, \sigma'}(t)$ according to~(\ref{ex rho path integral nonint}) (minding the matrix transposition). 

We immediately see that the resulting density matrix, calculated from the path integral on the KP contour, has the same form as that computed with ordinary quantum mechanics---see equations~(\ref{ex rho QM})---once the following identifications are made:   
\begin{eqnarray}
n_{\mathrm{f}, \uparrow} = W_{\uparrow, \downarrow} + W_{\uparrow} ~, \quad 
n_{\mathrm{f}, \downarrow} = W_{\uparrow, \downarrow} + W_{\downarrow} ~,
\label{equations for W}
\end{eqnarray}
where the weights $W$ in the right-hand sides are those appearing in~(\ref{ex rho QM}). One solution of equations~(\ref{equations for W}) is given by the grand-canonical statistical weights
\begin{eqnarray}
  W_{\uparrow} = \mathcal{Z}^{-1}  \mathrm{e}^{- \beta \left(\Delta - \mu \right)} ~,  \quad 
W_{\downarrow} = \mathcal{Z}^{-1}  \mathrm{e}^{- \beta \left(- \Delta - \mu \right)} ~, \quad 
W_{\uparrow \downarrow} = \mathcal{Z}^{-1}  \mathrm{e}^{2 \beta \mu } ~, 
\end{eqnarray}
where $\mathcal{Z}$ is the equilibrium partition function given in~(\ref{Z equilibrium spin problem}). Note that $\Delta \neq 0$ (i.e.~a non-thermal choice for $\hat{\cal H}_{\mathrm M}$) is essential to realize spin symmetry breaking in our example, as it allows for $W_{\downarrow} \neq W_{\uparrow}$ and to tune the relative weights at will. Moreover, taking the limit $\Delta \rightarrow + \infty$ allows to prepare the system in the $\left| \downarrow \right\rangle$ state ($W_{\downarrow} = 1$, $W_{\uparrow} = 0$) and, viceversa, taking the limit $\Delta \rightarrow - \infty$ allows to prepare the system in the $\left| \uparrow \right\rangle$ state ($W_{\downarrow} = 0, W_{\uparrow} = 1$). 

This simple example illustrates the power of the path integral approach on the KP contour, which gives us the full freedom of defining the Hamiltonian on the Matsubara branch in an arbitrary way.

\section{Summary and conclusions}
\label{sect:summary}

In summary, we have presented a discrete temporal procedure for the rigorous construction of the nonequilibrium path integral on the KP time contour. Our main result is that we have rigorously converted the general (contour-independent) expressions given by equations~(\ref{S_f}) and~(\ref{S_b}) into the forms of equations~(\ref{fermions continuum}) and~(\ref{bosons continuum}), via the expression of the GF, equation~(\ref{G discrete continuum limit}), that is specific of the KP contour.

Our procedure generalizes the one used for the SK contour~\cite{Kamenev}, allowing us to include the imaginary-time (Matsubara) branch in addition to the real-time branches. Consequently, path-integral theories on the KP time contour can account, in principle, for arbitrarily correlated initial states and/or statistical mixtures, or particular chosen pure states. 

Of course, many of the results that can be usually achieved with path integrals can also be obtained by applying other methods, such as the equations of motion for nonequilibrium GFs. However, path integrals are extremely useful when part of the fields entering the action can be integrated away, and/or when a suitable Hubbard-Stratonovich transformation on the interaction term yields a meaningful physical theory which can then be studied, e.g., via a saddle-point approximation. Interested readers can find the application of all these procedures in the context of a theory of laser-induced nonequilibrium superconductivity~\cite{Secchi17}.

\section*{Acknowledgements}

This work was supported by the European Union's Horizon 2020 research and innovation programme under grant agreement No.~785219 - GrapheneCore2. We thank Alex Kamenev for very stimulating comments on our manuscript.

\end{document}